# Spatial navigation in preclinical Alzheimer's disease: A review

Syrine Salouhou[1], Victor Gilles[2], Remi Vallée[1], Gillian T. Coughlan[3], Romain Bachelet[2], Michael Hornberger[4], Hugo Spiers[5], Antoine Coutrot[1]*and Antoine Garnier-Crussard[2,6†]*

[1] CNRS, INSA Lyon, Université Claude Bernard Lyon 1, LIRIS, UMR 5205, Villeurbanne, France
[2] Clinical and Research Memory Centre of Lyon, Lyon Institute for Aging, Charpennes Hospital, Clinical Research Center Ageing-Brain-Frailty, Hospices Civils de Lyon, Villeurbanne, France
[3] Department of Neurology, Massachusetts General Hospital, Harvard Medical School, Boston, MA, USA,
[4] Faculty of Medicine, University of Southampton UK, Southampton, United Kingdom
[5] Department of Experimental Psychology University College London, London, United Kingdom
[6] University Claude Bernard Lyon 1, CNRS, INSERM, CRNL, UMR5292, Bron, France

[†]e-mail: antoine.garnier-crussard@chu-lyon.fr
*co-senior authors

**Abstract |** Alzheimer's disease (AD) develops over a prolonged preclinical phase, during which neuropathological changes accumulate long before cognitive symptoms emerge. Identifying cognitive functions affected at early stages is critical for the preclinical detection of asymptomatic individuals at-risk of AD, as early risk identification could enable timely interventions aimed at mitigating the development of future significant cognitive impairment. While episodic memory decline typically appears after substantial medial temporal lobe damage, spatial navigation has emerged as a particularly sensitive cognitive function in preclinical AD. In this review, we provide an overview of spatial navigation computations and the tasks used to assess them, highlighting how spatial navigation relies on neural circuits corresponding to the earliest sites of AD pathology. We synthesize evidence from cognitively unimpaired individuals with AD biomarkers, i.e. individuals at-risk of AD and discuss future research directions. Overall, performance on spatial navigation tasks, particularly path integration and wayfinding, correlates with biomarkers of AD pathology. Spatial navigation assessment can represent a sensitive and scalable approach for early detection of individuals at-risk of AD in preclinical stages and will inform future interventions to mitigate the progression toward clinically significant cognitive impairment.

## Introduction

Alzheimer's disease (AD) represents one of the most pressing public health challenges of our time, with approximately 57 million people worldwide currently living with dementia, a number projected to triple by 2050[1]. Despite recent advances in disease-modifying treatments, the window for effective intervention remains narrow, and early detection tools that are scalable, accessible, and sensitive to the earliest cognitive changes are needed[2]. Although there have recently been dramatic advances in blood-based biomarkers for the early detection of AD pathology, their low positive predictive value (PPV) in cognitively healthy adults means they cannot be used as a standalone test in this healthy population [3,4].
Neuropathological changes in AD, including amyloid-beta deposition and tau tangle formation, begin 10 to 15 years before any observable clinical symptoms[5]. At this 'preclinical' stage, individuals show no noticeable cognitive impairment, yet very subtle changes may already be detectable[6]. Identifying such changes early offers an opportunity to anticipate the disease course, stratify populations for clinical trials, and initiate preventive or therapeutic strategies before irreversible neuronal loss occurs. The terms ‘preclinical’ and ‘presymptomatic’ are subjects of debate within our community, given that not all patients at these stages go on to develop cognitive impairment during their lifetime[7,8]. In this review, we adopt the International Working Group (IWG) framework, which classifies preclinical stages as ‘asymptomatic at risk’ or ‘presymptomatic’ according to biomarkers profiles and genetic risk in individuals who carry pathophysiological biomarkers of AD pathology (amyloid, tau) without cognitive symptoms[5,9].
Among candidate early markers, spatial navigation has emerged as particularly promising. Spatial navigation relies on brain regions affected in the earliest phases of AD, including the entorhinal cortex, hippocampus, and retrosplenial cortex, and is thus theoretically vulnerable before symptoms of episodic memory decline become measurable[10]. Crucially, the rapid development of digital assessment tools, including tablet, mobile, and virtual reality platforms, now makes large-scale, standardized navigation testing feasible outside of specialized memory clinics[11,12].
While previous reviews have established the neurobiological rationale for studying spatial navigation in AD, and meta-analyses have documented its decline across symptomatic stages, i.e. mild cognitive impairment (MCI) or dementia[13,14], no review to our knowledge has specifically focused on the preclinical stage, nor synthesized the rapidly expanding evidence linking experimental navigation tasks to in vivo AD biomarkers in cognitively unimpaired individuals. This constitutes the specific and novel contribution of the present review.
We will first summarize the early pathophysiological changes in AD that disrupt the neural circuits underlying spatial navigation. We then provide an overview of the experimental tasks used to assess navigation, before synthesizing existing findings on navigation performance in preclinical populations and their relationship to AD biomarkers.

## Preclinical Detection of Alzheimer's Disease: The Potential of Spatial Navigation

### The diagnostic landscape of AD and the role of cognitive assessment

AD is typically diagnosed at a moderate stage, when symptoms are already associated with significant functional decline[15]. Current diagnostic procedures combine neuropsychological testing, structural neuroimaging, and pathophysiological biomarkers from cerebrospinal fluid or PET imaging. These procedures are time-consuming, expensive, and largely restricted to specialized centers, contributing to diagnostic delays and geographic inequalities in access to care[1,16].
Blood-based biomarkers represent a major recent breakthrough, increasingly positioned as frontline tools in both primary and secondary care[17–19]. The revised Alzheimer’s Association ( AA) criteria now rely exclusively on a biological definition of AD[20], while the International Working Group (IWG) frameworks maintain a combined biological and clinical definition[5,9]. In this context, cognitive assessments serve as essential complements: they help reduce false-positive rates by situating biomarker results within a broader clinical picture and provide sensitive outcome measures for tracking disease progression and treatment response[4]. However, widely used screening tools such as the Mini-Mental State Examination lack the sensitivity required to detect the earliest cognitive changes of AD and are strongly influenced by non-pathological factors such as education level[21,22].

**Beyond episodic memory: the case for spatial navigation as an early marker**

Episodic memory deficits have traditionally been regarded as the hallmark of early AD[5]. However, such impairments only become measurable after substantial medial temporal lobe pathology has already accumulated, and they are not specific to AD[10,23,24]. This has motivated the search for earlier and more specific cognitive markers.

Spatial navigation, the ability to identify, plan, and maintain a route toward a destination, is a fundamental cognitive function that underpins autonomy, safety, and quality of life in older adults[25]. Crucially, it depends on the entorhinal cortex, hippocampus, and retrosplenial cortex, which are among the first regions affected by AD pathology, making navigation theoretically vulnerable at stages that precede episodic memory failure. Compared to classical cognitive assessments, spatial navigation measures are less influenced by verbal and educational factors [26], and may offer a more standardized, internationally applicable means of identifying early cognitive signatures of AD.

Meta-analyses and literature reviews have consistently documented that spatial navigation skills decline progressively throughout the course of AD, with performance measures showing large effect sizes in distinguishing healthy older adults from individuals with MCI, particularly amnestic MCI, and AD dementia[13]. Research findings increasingly converge on spatial navigation as one of the earliest cognitive functions to be affected[10,27,28].

Beyond diagnostics, spatial navigation assessments may serve as functional outcome measures reflecting real-world difficulties as symptoms emerge, and potentially as sensitive indicators of treatment response in clinical trials. The growing development of disease-modifying treatments, including monoclonal antibodies targeting amyloid pathology[29,30], underscores the need for early diagnosis, as these therapies appear most effective when initiated early in the disease trajectory[31].

The development of digital navigation tools, including tablet-based, mobile, and virtual reality platforms, has significantly expanded the feasibility of large-scale assessment [32]. Tools such as Sea Hero Quest have demonstrated that ecologically valid navigation data can be collected across diverse populations, enabling normative benchmarking and broadening access beyond specialized clinical settings[11,12,33].

**Search strategy and study selection**

A systematic literature search was conducted on PubMed, Web of Science, and Google Scholar using a combination of keywords targeting spatial navigation, AD, preclinical stages, relevant biomarkers, and autosomal-dominant AD genes. Spatial navigation terms included: navigat*, allocentric, egocentric, wayfinding, and path integration. Disease terms included: Alzheimer* and AD. Population and stage terms included: preclinical, presymptomatic, subjective cognitive decline (SCD), isolated cognitive complaints (ICC), and PSEN1/PSEN2/APP. Biomarker terms included: amyloid beta, tau, phosphorylated tau, NfL. Search syntax was adapted to the conventions of each database.

Studies were included if they assessed spatial navigation using experimental tasks in cognitively unimpaired individuals with plasma, CSF or PET AD biomarkers. Studies restricted to MCI participants, or defining at-risk groups solely on the basis of APOE genotype or family history via questionnaire, were excluded. Although studies on autosomal-dominant AD would have been relevant, none meeting our inclusion criteria were identified.

## Spatial Navigation Computation and Early AD Pathology

Spatial navigation is a remarkably complex cognitive function. It depends on the integration of both self-motion signals (derived from one's own movement) and environmental cues (stable features of the surrounding environment)[34].

These sources of information give rise to two complementary spatial reference systems that support the construction of lasting internal representations of the environment: an egocentric system (centered on the self) and an allocentric system (centered on the external world)[35].

These complementary egocentric and allocentric systems support two key mechanisms for navigation: path integration and cognitive mapping. Path integration uses self-motion cues, including vestibular and proprioceptive signals, motor efference copies, and optic flow, to continuously track one's displacement from a starting point[36,37]. This system can operate independently of landmarks, though environmental cues can recalibrate position and heading when available[34]. In parallel, the cognitive map represents a mental model of the environment, encoding spatial relationships among landmarks and supporting flexible navigation strategies, such as planning novel routes or inferring shortcuts. Together, these mechanisms allow the brain to integrate internal motion signals and external environmental information, providing a robust framework for spatial orientation and movement within complex environments[38]. Importantly, in AD, these same circuits are among the first to be affected both by amyloid deposition and tau pathology. We will focus on early tau pathology (leading to neurofibrillary tangles) as it may specifically disrupt the neural computations underlying spatial navigation.

Neurofibrillary tangles first appear in the most superficial layer of the trans-entorhinal cortex and in subcortical nuclei (Braak stage I) before extending to the entorhinal cortex and Ammon's horn of the hippocampus (Braak stage II)[39]. These early-affected regions are central hubs for spatial navigation processing.

Within this network, navigation relies on the activity of specialized neurons: place cells in the hippocampus encode specific locations, grid cells in the medial entorhinal cortex fire in a regular hexagonal pattern across the environment, and border and boundary cells in the entorhinal cortex and subiculum respond to environmental limits[38,40,41]. Speed cells in the medial entorhinal cortex further contribute by encoding the animal's movement velocity[42]. Importantly, the entorhinal and trans-entorhinal cortices also receive strong vestibular inputs, which are crucial for maintaining the stability of place and head-direction cells[43]. Consistent with this, neuroimaging studies show that reduced vestibular function, including saccular impairment, is associated with morphometric alterations in these cortices[44–46]. Thus, the first regions affected by tau pathology overlap with those supporting the neural computations essential for spatial navigation.

In Braak stages III–IV, tau pathology extends from the medial temporal lobe to limbic and associative cortices. Aggregates accumulate within lesions from the II stages, before spreading to the neocortex of the fusiform and lingual gyri (Braak stage III). Pathology then progresses to the high-order association cortices in the prefrontal, parietal, and temporal lobes (Braak stage IV). The retrosplenial cortex (RSC), which becomes functionally implicated during these stages, plays a central role in spatial navigation. Strongly interconnected with the entorhinal cortex[47], the RSC contains head direction cells and participates in the integration of egocentric and allocentric spatial frames of reference[48]. These interactions are thought to mediate the communication between parietal egocentric and medial temporal allocentric systems[49,50]. Note that at these stages, cognitive symptoms (beyond spatial navigation) appear, and individuals enter the clinical stage of AD (MCI or dementia)[51].

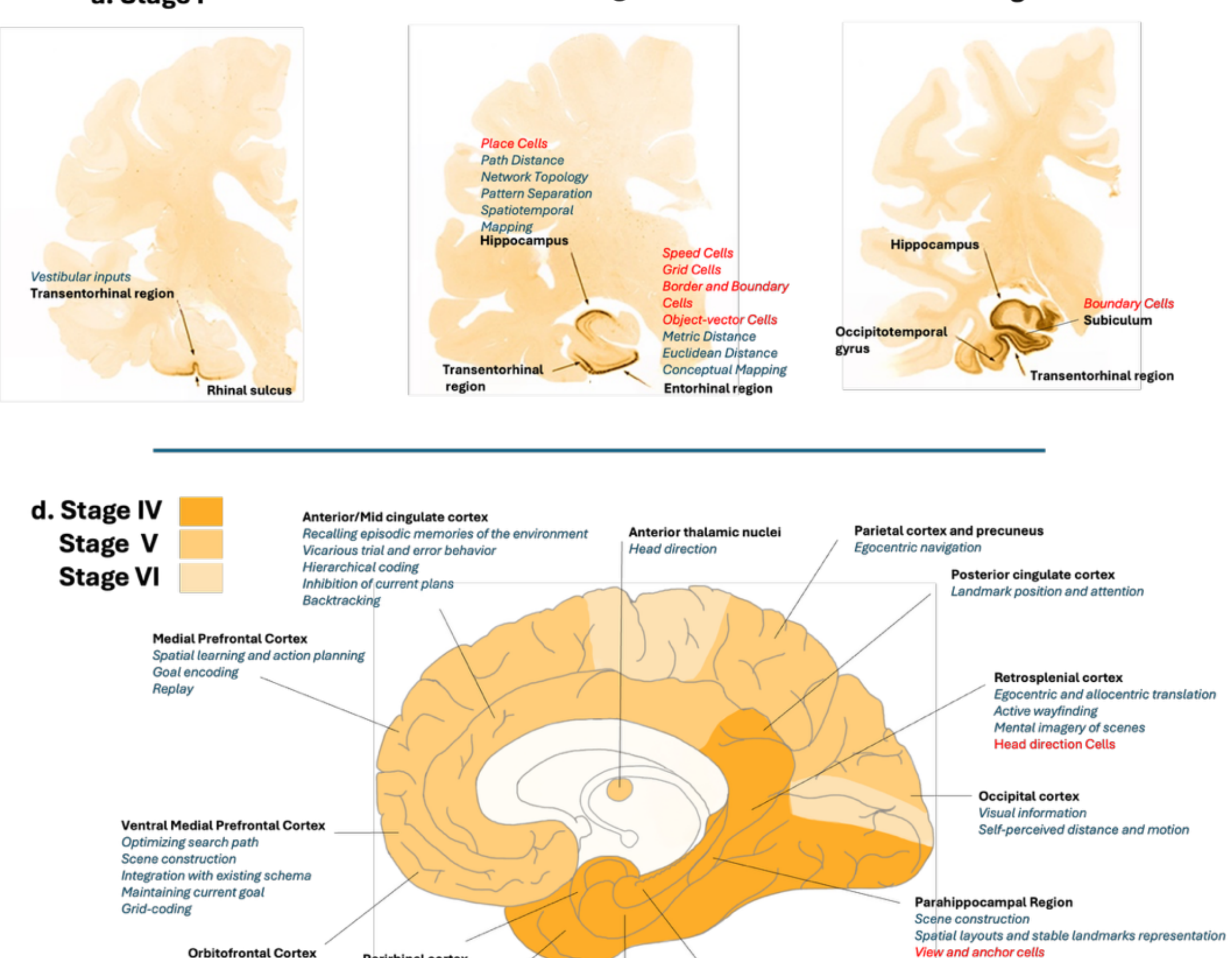

Figure 1 | **Cognitive processes and cell types involved in spatial navigation: anatomical overlapping with Braak stages of Alzheimer's disease. a–c,** Progression of tau pathology (black shading represents AT8-immunoreactive tau pathology) across the first three Braak stages. **d,** Braak stages IV–VI: stage IV is shown in dark yellow, stage V in light yellow, and stage VI in beige. Across all panels, blue labels indicate cognitive processing involved in spatial navigation, whereas red labels denote cell types known to encode spatial navigation. Panels a-c are adapted from Braak et al. 2006 and panel d is inspired from Coughlan et al. 2018. Braak stages were derived from Tau-PET imaging based on Malpetti & La Joie 2022, not from neuropathological studies.

The hippocampus and entorhinal cortex construct spatial representations resembling internal maps, while posterior regions such as the parahippocampal and retrosplenial cortices provide contextual cues that anchor and stabilize these maps with respect to external landmarks. These spatial representations are subsequently integrated with frontal circuits to support goal-directed route planning and navigation[38].
Ultimately, tau aggregates extend into neighboring neocortical regions, resulting in widespread cortical atrophy (Braak stages V–VI). The pathology subsequently progresses to first-order sensory association and premotor areas (stage V) and eventually reaches the primary neocortical fields (stage VI).

## Spatial Navigation Framework and Experimental Designs

Several factors influence how cognition is engaged during navigation. These include the spatial scale of the environment (for example, navigating a small room versus a large city) [53,54], the availability and reliability of environmental cues[55–57], and the navigator's prior knowledge or familiarity with the space[58,59], including awareness of starting points and intended destinations[34]. Moreover, individual and socio-demographic factors can also modulate navigation performance[60,61,62]. Therefore, there are several ways to assess spatial navigation. Everyday navigation represents a highly integrative behavior that draws on perceptual, memory, and executive systems[14,63–65]. It involves combining multiple types of spatial information, selecting context-appropriate strategies, and flexibly switching between them when circumstances change. Moreover, effective navigation requires the coordination of numerous sub-processes, making it one of the most cognitively demanding and multifaceted human abilities [34,66–68].

In the context of evaluating spatial navigation as a screening tool for AD, several different types of tasks have been used, including path integration tasks, cognitive mapping tasks, wayfinding tasks, GPS-based driving measures, static visuospatial tasks, and spatial navigation questionnaires.

### Assessing Spatial Navigation: Task overview

Path integration tasks require participants to follow a predefined path in three dimensions environment and then return to their starting location, typically without any external landmarks or visual cues. Various performance metrics can be derived from these tasks, such as rotation error, distance error, or drop error. Several variants of path integration tasks exist, for instance, those including or excluding landmark support[70,73,74,76].

Another approach is the cognitive mapping task, in which participants are asked to locate landmarks on a 2D map after exploring an environment, either physically or virtually. Examples include the Super Market Task[69] (see Fig. 2).

Wayfinding tasks assess participants' ability to navigate toward a goal within a virtual or real-world environment[25,77,78]. Navigation may rely on a previously studied 2D map or on memory from self-guided exploration. Performance can be evaluated based on the accuracy and efficiency with which participants reach specific landmarks or target locations such as Sea Hero Quest, the Navigation Test Suite[74], the Floor Maze Task[72] or the Barn Ruins [75] (see Fig. 2). Importantly, performance on some navigation tasks has been shown to predict real-world navigation behavior, supporting their ecological validity [79]. For instance, scores on Sea Hero Quest correlate with GPS-measured navigation during a similar task in the real world [11,80], and laboratory task performance has been shown to reflect real-world wayfinding difficulties in AD patients[81].

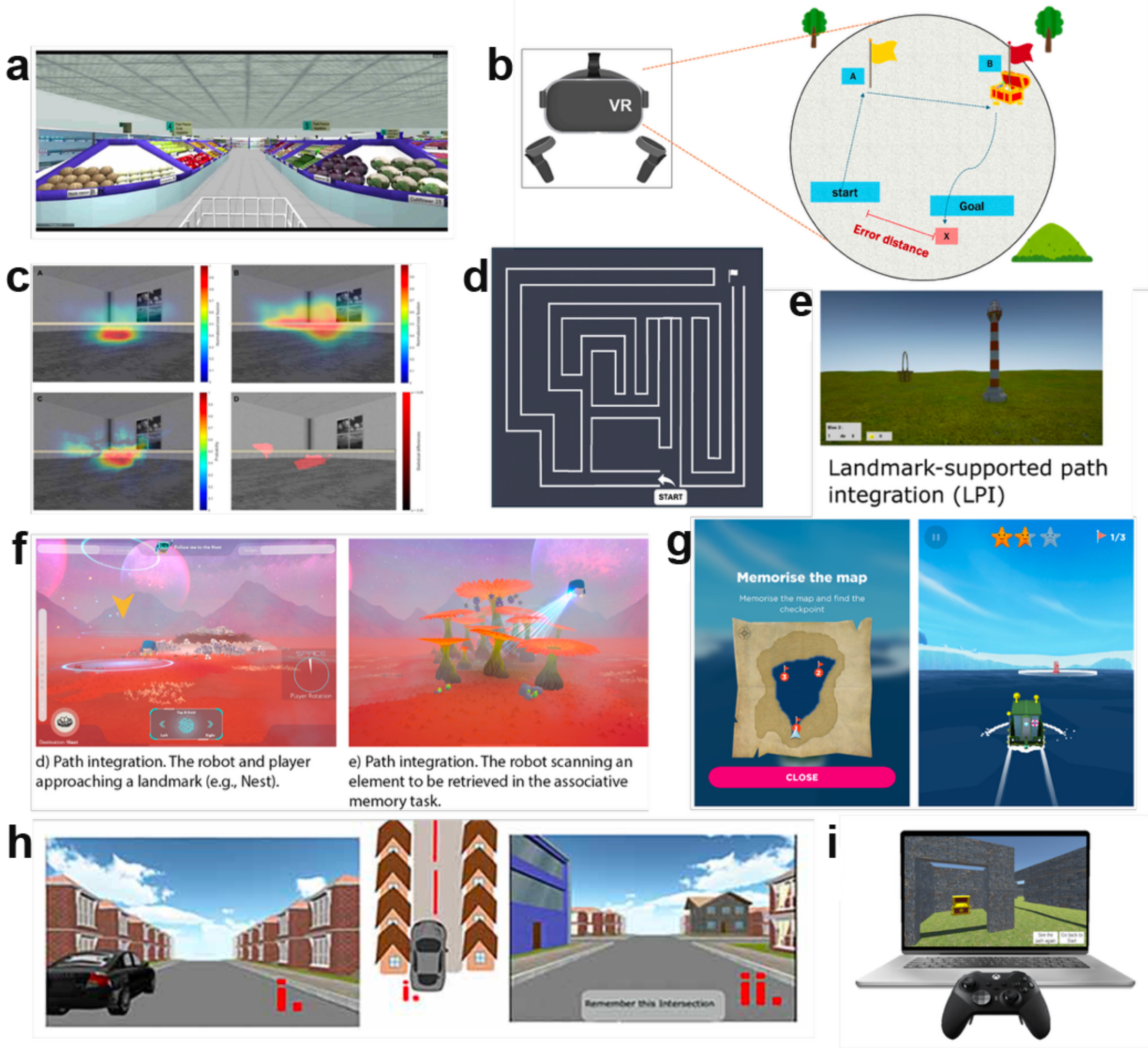

Figure 2 | **Spatial navigation tasks used in a preclinical context. a,** Super Market Task [69] **b,** 3D VR navigation task[70] **c,** Virtual Morris Water Navigation[71] **d,** Floor Maze Test [72] **e,** The Apple Game[73] **f,** Spatial Performance Assessment for Cognitive Evaluation (SPACE)[12] **g,** Sea Hero Quest[11] **h,** Navigation Test Suite[74] **i,** Barn Ruins[75]. Panel d has been adapted from the original published version with authors authorization.

GPS-based driving measures represent another way to assess spatial navigation abilities, by analyzing real-world driving behavior through GPS tracking, focusing on various navigation-related metrics. For example, Ghosh et al. (2022) adopt a mechanistic, fine-grained approach using machine learning applied to features from complementary domains, including mobility and spatial entropy measures, trajectory shape descriptors, and graph-theoretical topology analyses, enabling the detection of subtle alterations in navigation patterns.
Static visuospatial tasks, such as the Four Mountains Task[10,83] or the Visual Perspective Taking Task[84], assess spatial processing in static conditions, without active movement through space.

Finally, spatial navigation questionnaires, such as the Santa Barbara Sense of Direction Scale[84], can be administered to participants or close informants to assess perceived navigation abilities and everyday orientation performance.

After summarizing the different approaches used to estimate and measure spatial navigation, we now describe studies assessing the associations between spatial navigation and Alzheimer's disease pathology measured in vivo in cognitively healthy individuals.

## Evidence Linking Spatial Navigation Deficits to Preclinical AD

Several studies have evaluated the diagnostic accuracy of spatial navigation measures in preclinical AD, with ROC-based analyses providing the clearest estimates of classification accuracy. Among the most robust findings, Shima et al. (2025) showed that path integration (PI) errors strongly discriminated plasma p-tau181 levels (AUC = 0.86, sensitivity = 91.7%, specificity = 77.8%), with PI errors also correlating with GFAP, and NfL. Allison et al. (2019) similarly reported AUC values ranging from approximately 0.72 to 0.77 across several components of a cognitive mapping task, which corresponds to a moderate (or fair) level of discrimination according to commonly used interpretations of ROC curve analyses, with landmark identification emerging as the strongest discriminator of CSF Aβ42 status (AUC = 0.77).

Table 1 | **Spatial navigation assessments in cognitively healthy individuals with Alzheimer's diseases biomarkers.**

| **Task Name** | **Spatial Navigation Task (Type)** | **Participants (n) Age, mean (years)** | **Biomarkers or AD conversion** | **Task measure** | **Regression analysis** | **ROC analysis** |
|---|---|---|---|---|---|---|
| ***3D Navigation*** | | | | | | |
| Floor Maze Test<br>(Tangen et al. 2022) | Wayfinding<br>*Real Space* | n=332;<br>Age=69.9 | 2-year AD Conversion | Total Time | OR = 1.07<br>$p = 0.06$ | AUC: 0.75<br>Se: 0.70<br>Sp: 0.70 |
| | | | CSF P-tau | Total Time | $\beta = 0.12$<br>$p = 0.03$ | NA |
| | | | CSF NfL | Total Time | $\beta = 0.13$<br>$p = 0.02$ | NA |
| | | | CSF Aβ42/Aβ40 | Total Time | $\beta = 0.04$<br>$p = 0.49$ | NA |
| Cognitive Mapping task<br>(Allison et al. 2019) | Cognitive Mapping<br>*Desktop Computer* | n=30; Age=73±4 | CSF Aβ42 | Landmarks identification | $\beta = 0.664$<br>$p = 0.003$ | AUC: 0.77<br>Se: 0.80<br>Sp: 0.67 |
| Path Integration Task<br>(Shima et al. 2025) | Path Integration<br>*Virtual Reality* | n = 111 ; Age = 54.8 ± 12.2 | Plasma p-tau181 | Error distance | $r = 0.327$<br>$p < 0.001$ | AUC: 0.86<br>Se: 0.92<br>Sp: 0.78 |
| | | | Plasma GFAP | Errors distance | $t = 2.18$<br>$p = 0.03$ | NA |
| The Apple Game<br>(Colmant et al. 2025) | Pure Path Integration<br>*Desktop Computer* | n = 75; Age = 70 | Tau PET in the MTL | Rotation error | Estimate =33.2<br>$t = 3.56$<br>$p < 0.001$ | NA |
| | | | Amyloid status (CSF/PET) | Drop error | $t = 2.29$<br>$p = 0.02$ | NA |
| Supermarket Task<br>(Coughlan et al. 2023) | Cognitive Mapping<br>*Tablet* | n = 1875; Age = 65.9±7.5 | CSF p-tau181 /Aβ1−42 | SMT score. | $\beta = -0.04$<br>$p = 0.001$ | NA |
| | | | CSF Aβ1−42 | Task errors. | $\beta = 0.03$; $p = 0.04$<br><br>$p = 0.058$.<br>adjusting for APOE | NA |
| Morris Water Maze<br>(Chen 2024) | Human analog of the Morris Water Maze Test<br>*Real space* | n = 61 | Plasma p-tau217, Plasma p-tau181 | Navigation errors | *p-tau217 :*<br>$r = 0.34$, $p = 0.003$<br><br>*p-tau181 :*<br>$r = 0.34$, $p = 0.003$ | NA |
| ***Questionnaires*** | | | | | | |
| Self-Report Questionnaire<br>(Allison et al. 2019) | Self-Reported change in spatial navigation ability<br>*Questionnaire* | n = 30; Age = 73±4 | CSF Aβ42 | Reported navigation ability | $\beta = -0.447$<br>$p = 0.016$ | AUC: 0.72<br>Se: 0.87;<br>Sp: 0.67 |
| ECog<br>(Levine et al. 2024) | Self and informant-reported navigation<br>*Questionnaire* | n ≈ 370; Age = 73 ±7 | CSF ptau181/Aβ42 | Reported navigation ability | *Self-reported*<br>$\beta = 0.059$<br>$p = 0.59$<br><br>*Informant-reported*<br>$\beta = 0.104$<br>$p = 0.36$ | NA |
| Santa Barabara Sense of Direction Scale<br>(Levine et al., 2022) | Self-Reported Spatial Ability (longitudinal)<br>*Questionnaire* | n = 180; Age = 72 ±5 | CSF Aβ42, p-tau181 and p-tau181/Aβ42 | Reported Navigation Ability | CSF Aβ42 :<br>$\beta < 0.001$<br>$p = 0.16$<br><br>CSF p-tau181:<br>$\beta < -0.001$<br>$p = 0.49$<br><br>CSF p-tau181/Aβ42 ratio:<br>$\beta = -0.239$<br>$p = 0.17$ | NA |
| ***Static visuo spatial Assesment*** | | | | | | |

| 4 Mountains Task (Coughlan et al. 2023) | Allocentric Translation<br>*Tablet* | n = 1875; Age = 66±7 | CSF p-tau181 | Task Errors | β = −0.04<br>p = 0.0015 | NA |
|---|---|---|---|---|---|---|
| 4 Mountains Task (Lorentzen et al. 2025) | Allocentric Spatial Working Memory<br>*Tablet/Computer* | n = 61; Age = 68.3 | CSF Aβ Status<br>CSF p-tau181 | Task Errors | *Aβ+* : β =1.13, p = 0.09<br>*Tau+* : β = 0.53, p = 0.47 | NA |

AD, Alzheimer's disease; Aβ, amyloid-β; Aβ42/Aβ40, amyloid-β 42/40 ratio; Aβ+, amyloid-positive; Aβ−, amyloid-negative; APOE ε4, apolipoprotein E ε4 allele; AUC, area under the receiver operating characteristic curve; β, standardized regression coefficient; CI, confidence interval; CSF, cerebrospinal fluid; FMT, Floor Maze Test; GFAP, glial fibrillary acidic protein; MCI, mild cognitive impairment; MMSE, Mini-Mental State Examination; MTL, medial temporal lobe; NA, not applicable; NfL, neurofilament light chain; OR, odds ratio; PET, positron emission tomography; ROC, receiver operating characteristic; Se, Sensitivity; Sp, Specificity; SMT, Supermarket Task.

Tangen et al. (2022) found that the Floor Maze Test moderately predicted conversion to AD dementia at 2 and 4 years (AUC ≈ 0.74–0.75), with PI performance linked to CSF p-tau and NfL but not CSF Aβ42/40 ratios or APOE ε4.

Beyond ROC analyses, several studies have identified meaningful associations between behavioral navigation performance and AD biomarkers. Colmant et al. (2025) reported that rotation and drop errors in PI increased with medial temporal tau burden (Table 1), while Chen (2024)[76] found significant correlations between plasma p-tau217/p-tau181 and errors in a human analog of the Morris Water Maze (r ≈ 0.34). Coughlan et al. (2023) further demonstrated that higher CSF p-tau181 and lower Aβ1–42 predicted poorer performance on both their navigation task and the Four Mountains Test, although the magnitude of effect was small and with the p-tau181/Aβ1–42 ratio emerging as the strongest predictor.

However, the literature also documents numerous null or inconsistent findings, particularly for amyloid-related measures. Similarly, Lorentzen et al. (2025) found no biomarker-related differences on the Four Mountains Test, and several studies reported no associations between navigation performance and plasma markers such as NfL, GFAP, or Aβ42/40[70,76]. These discrepancies may reflect methodological variability or small effect sizes.

Results from spatial navigation questionnaires and static visual tasks are mixed. Levine et al. (2024) reported very weak or nonsignificant associations between self- or informant-rated spatial navigation questionnaires and AD biomarkers, whereas Allison et al. (2019) found moderate predictive value for CSF Aβ42 in self-informant questionnaire (AUC = 0.72, sensitivity = 0.87, specificity = 0.67). Similarly, egocentric tasks like the 4MT show inconsistent biomarker correlations[69] (see Table 1). Extended results are available in Table S1.

GPS-derived measures have shown excellent predictive performance. Bayat et al. (2021) reported very high accuracy (AUC = 0.96; 95% CI 0.90–0.98) using one year of continuous GPS data, with age and APOE ε4 status being the strongest predictors to differentiate amyloid-negative from amyloid-positive individuals. Similarly, Roe et al. (2023) and Chen & Babulal (2025) found that biomarker-positive individuals tended to reduce their driving frequency, trip distance, and route complexity. However, as noted by the authors, the strongest predictive metrics appear to reflect self-regulatory driving behaviors rather than pure spatial navigation ability, although the results remain highly promising for early detection of preclinical AD-related changes.

Despite generally positive trends, cross-study comparisons remain challenging due to substantial methodological heterogeneity, including differences in navigation paradigms, biomarker modalities, and cutoff values for biomarker positivity. These inconsistencies likely contribute to the variability in reported effect sizes.

Taken together, current evidence supports the potential of objective, well-characterized navigation tasks as indicators of early AD-related changes. Their clinical applicability would be strengthened by standardized task design and broader adoption of ROC-based analyses. It would also be useful to report negative and positive predictive values (NPV and PPV) according to the clinical context of use. Indeed, pre-test probability of having amyloid pathology considerably influences NPV and PPV[4]. Combined with accessible plasma biomarkers, navigation paradigms may offer a practical and informative approach for identifying preclinical AD in cognitively healthy individuals.

**Box 1 | From research to real-world benefit.** Spatial navigation assessment holds significant potential to complement existing diagnostic pipelines for AD, offering a low-cost, scalable, digitally deliverable tool for identifying subtle cognitive changes at the preclinical stage. If validated at the individual level, navigation testing could support earlier risk stratification, improve prevention trial outcomes, and broaden access to preclinical detection beyond specialized memory clinics.

Translational progress is encouraging. Several digital tools have already been deployed at scale across diverse populations, demonstrating feasibility in community and research settings. However, most evidence linking navigation performance to in vivo AD biomarkers still comes from small, selected cohorts, and no existing task yet achieves sufficient diagnostic accuracy for individual-level clinical use, with classification performances typically yielding moderate AUC values.

Several challenges must be addressed before clinical integration becomes realistic. Spatial navigation is cognitively complex, drawing on episodic memory, attention, executive function, and visuospatial abilities, meanings within a single task, different outcome measures or error types may reflect distinct underlying cognitive mechanisms[10]. Although navigation measures are generally less sensitive to verbal ability and cognitive reserve than conventional screening tools[26], they remain influenced by individual factors such as sex, cultural background, education, environmental exposure, and digital interfaces familiarity [61]. Large-scale data have also shown that country-level factors such as country of origin [60], socio-economic indicators [90] , or urban layout [91], can shape navigation performance, posing important challenges for normative modeling and cross-cultural comparability.

The path forward requires large longitudinal cohorts linking digital navigation performance to established AD biomarkers, harmonized assessment protocols for multicenter use, and robust normative databases stratified by demographics. Spatial navigation is unlikely to serve as a standalone marker, but as part of a multimodal cognitive and biological framework, it represents a promising and practically deployable component of future preclinical AD detection strategies.

Moreover, a broader investigation of cognitive functions sensitive to early AD pathological changes alongside spatial navigation could further improve our understanding of preclinical AD.

## Promising Research Avenue for Understanding Spatial Navigation in Preclinical AD

Several emerging research directions may further clarify spatial navigation alterations in preclinical AD. These include comorbidities, complementary physiological measures, sex- and hormone-related factors, socio-demographic factors, and early subcortical pathology, highlighting the multifactorial nature of navigation as an early AD marker.

Most studies on spatial navigation in AD have excluded significant comorbidities, despite their high prevalence, including hypertension, depression, and hypercholesterolemia[92]. Although comorbidities are known to negatively affect cognition across the AD spectrum[93,94], their specific impact on spatial navigation has been rarely investigated[95,96] and remains unexplored in preclinical AD. Addressing this gap may provide new insight into disease progression and individual prognosis [93]. Beyond general comorbidities, the integration of brain comorbidities (i.e., copathologies), notably TDP-43, Lewy bodies or vascular damages will be also interesting in future studies. Early work comparing spatial navigation in FTD and AD has shown distinct navigation profiles across these conditions[97], and studies in vascular cognitive impairment have similarly demonstrated navigation impairments linked to cerebrovascular pathology[98,99]. These findings highlight the potential of navigation tasks to differentiate AD from other etiologies of cognitive decline but also potential overlaps in spatial navigation deficits across pathologies.

Complementary physiological measures such as EEG and eye-tracking may provide additional insight into how spatial navigation is supported or altered in preclinical AD[100,101]. While behavioral tasks capture overall performance, EEG and gaze metrics can reveal early changes in visual exploration, attentional allocation, and neural processing. Evidence from later disease stages supports this possibility. For instance, Plaza-Rozales et al. (2023) showed that individuals with amnestic MCI who later progressed to AD dementia displayed both impaired navigation learning and altered gaze patterns, together with reduced occipital responses and disrupted beta-band connectivity during a virtual navigation task (see Fig. 2c). However, such approaches have not yet been applied to biomarker-positive cognitively healthy individuals, leaving their potential contribution to the characterization of preclinical navigation changes unexplored.

Women experience a higher incidence of AD than men, a difference that may in part reflect the neuroendocrine changes occurring across the menopausal transition[102]. Perimenopausal period constitutes a critical window of emerging vulnerability to AD-related brain changes where early intervention may be beneficial[103,104]. One study suggests that spatial navigation is modulated by hormonal status. Navigation strategy use appears sensitive to endocrine aging, with estrogen fluctuations influencing the balance between place- and response-based strategies[105]. Investigating spatial navigation performance during menopause may thus provide valuable insight into how hormonal transitions shape cognitive aging and early AD vulnerability.

Epidemiological data on spatial navigation in healthy populations remain scarce, limiting our understanding of interindividual differences, including potential sex effects. Sea Hero Quest, a mobile game played by over 3.7 million individuals worldwide, has begun to address this gap, revealing influences of age, gender, and cultural background on navigation performance[60,90]. These large-scale data enable the creation of population benchmarks, which can inform sensitive, easy-to-administer navigation tests and help detect early changes in at-risk groups, such as APOE*ε4 carriers[106,107].

GPS-based studies in cognitively unimpaired individuals with AD biomarkers have yielded promising results, particularly for identifying preclinical participants through driving behavior. These approaches are already valuable but extending them to include navigation-focused features could provide additional insights into spatial navigation strategies. For example, Ghosh et al. (2022) applied machine learning to a range of complementary navigation-related features and successfully distinguished AD patients from cognitively healthy controls. It would be interesting to explore whether such metrics are sensitive at the preclinical stage, particularly for differentiating cognitively healthy individuals with AD biomarkers from those without.

Early tau pathology affects subcortical nuclei, including the locus coeruleus (LC), one of the earliest regions involved in AD, which may contribute to subtle spatial navigation impairments at the preclinical stage[39]. By supporting contextual updating and flexible transitions between egocentric and allocentric navigation strategies, LC-related circuits play a role in spatial learning [108–111]. Early LC degeneration may therefore disrupt navigation independently of hippocampal atrophy or overt cognitive decline. Future work in humans could explore whether combining high-resolution imaging of the locus coeruleus with controlled navigation paradigms helps clarify links between LC integrity, navigation-related neural activity, and behavior.

## Conclusion

Spatial navigation is a promising cognitive domain for early detection of AD pathology, including at preclinical stages when individuals remain asymptomatic but already show biomarker evidence of pathology. The neural circuits supporting navigation, especially the entorhinal cortex, hippocampus, and retrosplenial cortex, are among the first affected by AD pathology, providing a strong biological rationale for assessing navigation deficits as early markers.

At present, tasks assessing path integration, wayfinding, and cognitive mapping appear particularly promising for detecting subtle changes in preclinical AD but are still far from implementation in daily practice. Performance on these tasks correlates with AD biomarkers.

While more studies are needed, spatial navigation assessment, particularly when combined with AD biomarkers, may transform preclinical screening, guide individualized care, and ultimately mitigate the impact of AD on patients, families, and healthcare systems. By moving beyond traditional memory-based assessments, spatial navigation offers a promising avenue to detect the earliest cognitive signatures of AD and clinical translation.

## Author contributions

S.S., A.C. and A.G-C. researched data for the article, made a substantial contribution to the discussion of content, and reviewed and edited the manuscript before submission. S.S. wrote the article. V.G., R.V., R.B., G.C., H.S., and M.H made a substantial contribution to discussion of content, and reviewed the manuscript before submission. A.C. and A.G-C. supervised this work and contributed equally.

## Competing interests statement

The authors Syrine Salouhou, Victor Gilles, Rémi Vallée, Romain Bachelet, Michael Hornberger, Hugo Spiers and Antoine Coutrot declare that they have no conflicts of interest. Independent of this work, Dr Antoine Garnier-Crussard is an unpaid sub-investigator or local principal investigator for clinical trials and studies sponsored by UCB Pharma, Biogen, TauRx Therapeutics, Roche, Novo Nordisk, Alzheon, Medesis Pharma, GlaxoSmithKline, AriBio, Acadia and received research grant from Roche, Lilly and Eisai. He declares that he has not received any personal funding and has not participated in any remunerated activities Dr Gillian Coughlan is also a consultant for Yale University.

## Funding

This work is supported by a grant from the French National Research Agency as part of the "Investissements d'Avenir ExcellencES" program from France 2030 (SHAPE-Med@Lyon ; ANR-22-EXES-0012) and by the ANR JCJC grant n°ANR-23-CE45-0023-01. Prof. Hugo Spiers is funded by a BIRAX grant (BIRAX\2 – 49BX21HSSA). Dr Gillian Coughlan is supported by a NIA-R00 (AG083063), an Alzheimer's Association Research Fellowship (23-1151259) and a Wellcome Trust CARE LEAP Program grant.

## Key points

- Spatial navigation deficits emerge at the preclinical stage of Alzheimer's disease, before episodic memory decline becomes measurable, offering a promising window for early detection.
- The brain regions first affected by AD pathophysiology, including the entorhinal cortex and hippocampus, are central nodes of the spatial navigation network.
- Cognitively unimpaired individuals with AD biomarker evidence show detectable spatial navigation impairments on experimental tasks, supporting its value as an early cognitive marker.
- Digital navigation tools enable ecologically valid, large-scale assessment of spatial cognition outside specialized clinical settings, broadening access to preclinical detection.
- Standardization of navigation tasks and integration with AD biomarker frameworks are needed to establish spatial navigation as a validated cognitive biomarker for preclinical AD.

## ToC blurb

*This Review synthesizes evidence linking experimental navigation tasks to in vivo AD biomarkers in cognitively unimpaired individuals, examines the neurobiological basis of early navigation deficits, and discusses digital assessment tools and their potential for scalable preclinical detection.*